\documentclass[showpacs,preprintnumbers,amsmath,aps,amssymb]{revtex4}
\usepackage{graphicx}
\usepackage[english]{babel}
\usepackage{bm}
\begin{document}
\baselineskip 16pt
\title{FRW in cosmological self-creation theory: Hamiltonian approach }
\author{Rafael Hern\'andez$^1$}
\email{fabulosorafa@hotmail.com}
\author{Juan M. Ram\'irez$^1$}
\email{jmramirz@fisica.ugto.mx}
\author{J. Socorro$^{1,2}$}
\email{socorro@fisica.ugto.mx}

\affiliation{$^1$Departamento de  F\'{\i}sica, DCeI, Universidad de
Guanajuato-Campus Le\'on,
 C.P. 37150, Le\'on, Guanajuato, M\'exico\\
  $^2$ Departamento de F\'isica, Universidad Aut\'onoma Metropolitana, Apartado Postal 55-534,
C.P. 09340 M\'exico, DF, M\'exico
}%

\date{\today}

\begin{abstract}
We use the Brans-Dicke theory from the framework of General
Relativity (Einstein frame), but now the total energy momentum
tensor fulfills the following condition
$\rm\left[\frac{1}{\phi}\left(8\pi T^{\mu \nu (M)}+T^{\mu\nu
(\phi)}\right)\right]_{;\nu}=0$. We take as a first model the flat
FRW metric in the Hamilton-Jacobi scheme and we present the
Lagrange-Charpit approach in order to find classical solutions. In
the quantum scheme, once we determine the characteristic surfaces,
the quantum solution is obtained. These two classes of solutions are
found for all values of the  barotropic parameter $\gamma$.
\end{abstract}

\maketitle
\title{}
\maketitle
\section{Introduction}
The Scalar-Tensor theories have their origin in the 50's. Pascual Jordan was intrigued by the appearance of a
new scalar field in Kaluza-Klein theories, especially in its possible role as a generalized gravitational
constant. In this way appeared the Brans-Dicke theory, with the particularity that each one of the energy momentum
tensor satisfy the covariant derivative \cite{Weinberg}, $\rm T^{\mu \nu (i)}\,_{;\nu}=0$, where $i$ corresponds to the
 i-th ingredient of matter content. A few years latter (1982), appeared a new proposal by Barber \cite{Barber1},
 known as self-creation cosmology (SCC) \cite{Barber,SSC10}. Since the original paper  appeared in 1982, more and
 more authors \cite{Singh,Singh0,Singh1,Singh2} have worked in the different versions of this theory in the classical
  fashion. By instant, Singh and Singh \cite{Singh3} have studied Raychaudhary-type equations for perfect fluid in
  self-creation theory. Pimentel \cite{Pimentel} and Soleng \cite{Soleng1,Soleng2} have studied in detail the
  cosmological solutions of Barber's self-creation theories. Reddy \cite{Reddy}, Venkateswarlu and
  Reddy \cite{Venka1}, Shri and Singh \cite{Shri1,Shri2}, Mohanty et al. \cite{Mohanty}, Pradhan and
  Vishwakarma \cite{Pradhan1,Pradhan2}, Sahu and Panigrahi
\cite{Sahu}, Venkateswarlu and Kumar \cite{Venka2} are some of the
authors who have studied various aspects of different cosmological
models in self-creation theory, many of them usually adopt a
particular ansatz for solve the Einstein field equations. These
papers adapted the Brans Dicke
 theory to create mass out of the universe's self contained scalar, gravitational and matter fields in simplest way.

The gravitational theory must be a metric theory, because this is the easiest way to introduce the Equivalence Principle.
However always it is possible to put additional terms to the metric tensor, the most obvious proposal is a scalar
field $\phi$. Recently Chirde and Rahate \cite{Rahate} investigated spatially homogeneous isotropic
Friedman-Robertson-Walker cosmological model with bulk viscosity and zero-mass scalar field in the framework of Barber's second
self-creation theory and found classical solutions, it is the simplest way to work with this theory because only
 takes in account the energy-momentum tensor of usual matter and the scalar field $\phi$.

This work is arranged as follow. In section II we present the method used in general way, where we take the Brans-Dicke Lagrangian density and we consider this in the self-creation theory. In section III, employing the flat FRW metric as a toy model with a barotropic perfect fluid, we found the energy density in
  dependence of the scalar factor and the scalar field. In Section IV we construct the Lagrangian and
   Hamiltonian densities for the cosmological model under consideration and it is presented the classical
    solution using the Hamilton-Jacobi approach. The classical solutions are obtained using the
    Lagrange-Charpit method \cite{Elsgoltz,delgado,lopez},
    which structure is such that we find classical solutions for all values of the $\gamma$ parameter.
     In this same section we solve the corresponding Wheeler-DeWitt equation for this case.
      The section V we present the conclusions of the work.

\section{Self Creation Cosmology in GR}

The Lagrangian density in the Brans-Dicke theory is
\begin{equation}
{\cal L}[g,\phi]= \sqrt{-g}\left(R\phi
-\frac{\omega}{\phi}g^{\mu\nu}\phi_{,\mu}\phi_{,\nu}\right)+\sqrt{-g}L_{matter},\label{L1}
\end{equation}
where $L_{matter}=16\pi\rho$, so making the corresponding variation to the scalar field and the tensor metric, the field equations in this theory become
\begin{equation}
R-\frac{\omega}{\phi^2} g^{\mu\nu}\phi_{,\nu}\phi_{,\mu}+\frac{2\omega}{\phi}\square^2\phi=0,\label{EC1}\\
\end{equation}

\begin{equation}
\rm
R^{\alpha\beta}-\frac{1}{2}g^{\alpha\beta}R=\frac{8\pi}{\phi}T^{\alpha\beta}
+\frac{\omega}{\phi^2}\left(\phi^{,\alpha}\phi^{,\beta}-\frac{1}{2}g^{\alpha\beta}\phi^{,\lambda}\phi_{,\lambda}\right)
+\frac{1}{\phi}\left(\phi^{,\alpha;\beta} -
g^{\alpha\beta}\square^{2}\phi \right),\label{ECG}
\end{equation}
where the left side is the Einstein tensor, the first term in the right side is the corresponding energy momentum tensor
 of material mass coupled with the scalar field $\phi^{-1}$. The second and third term corresponds at energy momentum
 tensor of the scalar field coupled also to $\phi^{-1}$. Both equations are recombined given the relation
\begin{equation}
 \rm \Box \phi=4\pi \lambda T, \label{dalamber}
\end{equation}
where $\lambda$ is a coupling constant to be determined from experiments. The measurements of the deflection of light
restrict the value of coupling to $|\lambda| < 10^{-1}$. In the limit $\lambda \to 0$, the Barber's second theory
approaches the standard general relativity theory in every aspect. $\Box \phi = \phi_{;\mu}^{;\mu}$ is the invariant
D'Alembertian and T is the trace of the energy momentum tensor that describes all non gravitational and non scalar field
matter and energy. Taking the trace of the equation (\ref{ECG}) and then substitute in (\ref{dalamber}), we obtain
the following wave equation to $\phi$
\begin{equation}\label{e:SSC1}
\Box\phi=\frac{8\pi}{(3+2\omega)}T_{M},
\end{equation}
comparing both equations (\ref{dalamber}) and (\ref{e:SSC1}), we
note that $\lambda=\frac{2}{(3+2\omega)}$ in the Brans-Dicke theory,
where $\omega$ is a coupling constant. In the Barber's theory, the
$\lambda$ parameter is a new coupling constant.

Now the Einstein equation can be rewritten as
\begin{equation}
\rm G^{\alpha \beta}=\frac{8\pi}{\phi}T^{\alpha\beta (M)} +\frac{1}{\phi} T^{\alpha \beta (\phi)}=T^{\alpha \beta (T)},
\end{equation}
where
$$\rm T^{\alpha \beta (\phi)}=\frac{\omega}{\phi}\left(\phi^{,\alpha}\phi^{,\beta}-\frac{1}{2}\phi^{,\lambda}\phi_{,\lambda}g^{\alpha\beta}\right)
+\left(\phi^{,\alpha;\beta} - g^{\alpha\beta}\square^{2}\phi
\right).
$$
In Brans-Dicke theory, each one energy momentum tensor satisfy the covariant derivative, $\rm T^{\mu \nu}\,_{;\nu}=0$.
 In the self creation theory, we introduce that the total energy momentum tensor is which satisfy the covariant
 derivative, $\rm T^{\mu \nu (T)} \,_{;\nu}=0$, namely;
\begin{equation}\label{e:tensor}
\rm T^{\alpha \beta (T)} \,_{;\beta}=0, \qquad \Rightarrow \qquad
 \left[\frac{8\pi}{\phi}T^{\alpha\beta (M)} +\frac{1}{\phi} T^{\alpha \beta (\phi)}\right]_{;\beta}=0,
 \end{equation}
which imply that $\rm \left[ Q^{\alpha \beta (T)}
\right]_{;\beta}=\frac{\nabla_\beta \phi}{\phi} \left[ Q^{\alpha
\beta (T)} \right]$,  where $\rm Q^{\alpha \beta (T)}= \phi
T^{\alpha \beta (T)}$. This equation is the master equation which
gives the name of  self creation theory, because the covariant
derivative of this tensor have a source of the same tensor multiply
by a function of the scalar field $\rm \phi$.

\section{FRW in Self Creation theory}
We apply the formalism using the geometry of Friedmann-Robertson-Walker
\begin{equation}
 \rm  ds^2  =  - N^2(t) dt^2  + A^2(t)  \left[\frac{dr^2}{1-\kappa r^2} + r^2 d\theta ^2  + r^2 sen^2 (\theta )d\varphi ^2\right],
 \label{metric}
\end{equation}
where $N$ is the lapse function, $A$ is the scalar factor. Now we solve the equations (\ref{ECG}), (\ref{dalamber})
and (\ref{e:tensor}), with the aim to find solutions to density $\rho(t)$, scalar factor $A(t)$ and the scalar field $\phi(t)$.\\

Taking the transformation $\prime=\frac{d}{Ndt}=\frac{d}{d\tau}$ and using $T^{\alpha\beta}$ as a fluid perfect,
first compute the classical field equations (\ref{ECG}), together with the barotropic equation of state $P=\gamma \rho$,
 this equation become
\begin{equation}\label{e:ec3}
3\left(\frac{A'}{A}\right)^2+3\frac{A'}{A}\frac{\phi'}{\phi}
-\frac{\omega}{2}\left(\frac{\phi'}{\phi}\right)^2-8\pi\frac{\rho}{\phi}+3\frac{\kappa}{A^2}=0,
\end{equation}
\begin{equation}\label{e:ec4}
2\frac{A''}{A}+\frac{A'^2}{A^2}+2\frac{A'}{A}\frac{\phi'}{\phi}+\frac{\omega}{2}\left(\frac{\phi'}{\phi}\right)^2+\frac{\phi^{\prime
\prime}}{\phi} +8\pi\gamma\frac{\rho}{\phi}+\frac{\kappa}{A^2}=0,
\end{equation}
the equation (\ref{dalamber}) become
\begin{equation}
 3\frac{A^\prime}{A} \frac{\phi^\prime}{\phi}+ \frac{\phi^{\prime\prime}}{\phi} = 4 \pi \lambda (1-3\gamma)\frac{\rho}{\phi}.\label{dal}
\end{equation}
The covariant equation (\ref{e:tensor}) or conservation equation to
the total energy momentum take the following form
\begin{equation}\label{e:econserv}
\rm
3\frac{A''}{A}\frac{\phi'}{\phi}-3\omega\frac{A'}{A}\left(\frac{\phi'}{\phi}\right)^2
+\omega\left(\frac{\phi'}{\phi}\right)^3-3\frac{A'}{A}\left(\frac{\phi'}{\phi}\right)^2
-24\pi\frac{A'}{A}(1+\gamma)\frac{\rho}{\phi}-\omega\frac{\phi''}{\phi}\frac{\phi'}{\phi}+
8\pi\frac{\phi'}{\phi}\frac{\rho}{\phi}-8\pi\frac{\rho'}{\phi} = 0,
\end{equation}
thus, the system equations to be solved are (\ref{e:ec3})-(\ref{e:econserv}).\\

We write the term $\left(\frac{ A^\prime}{A}\right)^2+\frac{\kappa}{A^2}$, in equation (\ref{e:ec3}), using equations (\ref{e:ec4}) and (\ref{dal}), and after
some algebra we have
\begin{equation}
\rm 3\frac{A^{\prime\prime}}{A}-3\frac{A\prime}{A} \frac{\phi^\prime}{\phi}+\omega\left(\frac{ \phi^\prime}{\phi} \right)^2 =
 2\pi\left[3\lambda(3\gamma-1)-2(1+3\gamma)\right]\frac{\rho}{\phi}. \label{equ}
 \end{equation}
Equation (\ref{e:econserv}) can be rewritten as
\begin{equation}\label{e:covar}
\rm\left(3\frac{A''}{A}-3\frac{A'}{A}\frac{\phi'}{\phi}+\omega\left(\frac{\phi'}{\phi}\right)^2\right)
\frac{\phi'}{\phi}-\left(3\frac{A'}{A}\frac{\phi'}{\phi}+\frac{\phi''}{\phi}\right)\frac{\phi'}{\phi}\omega
-24\pi(1+\gamma)\frac{\rho}{\phi}\frac{A'}{A}-8\pi\left(\frac{\rho}{\phi}\right)'=0,
\end{equation}
and using the equations (\ref{dal}) and (\ref{equ}) in (\ref{e:covar}), then we have the master equation for solve the energy density of the model, as
\begin{equation}
\rm
2\left[\lambda(3\gamma-1)(3+2\omega)-2(1+3\gamma)\right]\frac{\rho}{\phi}\frac{\phi'}{\phi}
-24(1+\gamma)\frac{\rho}{\phi}\frac{A'}{A}-8\left(\frac{\rho}{\phi}\right)'=0,
\end{equation}
defining the function $\rm F=\frac{\rho}{\phi}$, we have
\begin{equation*}
 \rm \frac{d}{d\tau} Ln \left[F A^{3(1+\gamma)}
 \phi^{\frac{1}{4}[\lambda(3\gamma-1)(2\omega+3)-2(1+3\gamma)]}\right]=0,
\end{equation*}
 who solution is
\begin{equation}
 \rm \rho=M_\gamma  A^{-3(1+\gamma)} \phi^\beta, \qquad \beta=\frac{(1-3\gamma)}{4}\left[2-\lambda a_0
 \right], \qquad a_0=3+2\omega.
\label{rho}
\end{equation}
This equation is equivalent to General Relativity expression \cite{Barber} with the addition the last factor representing
the self creation cosmology. Note that for a photon gas $\gamma=\frac{1}{3}$, so $\beta=0$ and equation (\ref{rho}) reduce
 to its General Relativity expression $\rho=\rho_{0}(A/A_{0})^{-4}$ which is consistent, because in the radiation epoch
 there was not interaction between photon and the scalar field.

\section{Lagrangian and Hamiltonian densities in SCC}
We will use the classical Hamilton-Jacobi approach to find solutions to $(\rho,A,\phi)$. The corresponding Lagrangian
 density using (\ref{metric}) into (\ref{L1})
\begin{equation}
\rm{\cal L}= 6\frac{A \phi {\dot A}^2 }{N}+ 6\frac{A^2 \dot A \dot
\phi}{N} -\frac{\omega A^3{\dot \phi}^2}{N \phi}+16\pi A^3 \rho N,
\end{equation}
the momenta are $(\Pi_{q}=\frac{\partial \cal L}{\partial\dot{q}})$
\begin{eqnarray}
&&\rm \Pi_A=\rm 12 \frac{A \phi \dot A}{N}+6\frac{A^2 \dot \phi}{N}, \qquad \Pi_\phi=\rm 6\frac{A^2 \dot A}{N}- 2\frac{\omega A^3 \dot \phi}{N \phi},\nonumber\\
&&\rm \dot A = \frac{N}{6(3+2\omega) \phi A^2}
\left( 3\phi \Pi_\phi+\omega A \Pi_A\right), \label{pia}\\
&&\rm \dot \phi = \frac{N}{2(3+2\omega) A^3}\left(A \Pi_A - 2 \phi
\Pi_\phi \right), \label{piphi}
\end{eqnarray}
when we write the canonical Lagrangian density $\rm {\cal L}_{canonical}=\Pi_q \dot q- N{\cal H}$, we obtain the corresponding Hamiltonian density as
\begin{equation}
\rm {\cal H}= \frac{A^{-3}}{12a_0 \phi}[-6\phi^2 \Pi_\phi^2 +\omega
A^2 \Pi_A^2 +6A \phi \Pi_A \Pi_\phi
 -192 \pi a_0 A^6 \phi \rho], \qquad a_0=(3+2\omega). \qquad  \label{hami1}
\end{equation}

\subsection{Classical scheme: Hamilton-Jacobi equation}
Using the gauge $\rm N= 12a_0 \phi A^3$, and the transformation in the momenta $\rm \Pi_q=\frac{\partial S}{\partial q}$, where
 S is known as the superpotential function, with this, the Hamiltonian density is written as (we include the equation (\ref{rho}) in this last equation)
\begin{equation}
\rm -6\phi^2 \left(\frac{\partial S}{\partial \phi}\right)^2 +
\omega A^2 \left(\frac{\partial S}{\partial A}\right)^2
 +6A \phi \left(\frac{\partial S}{\partial A}\right) \left(\frac{\partial S}{\partial \phi}\right)
 - \eta_\gamma A^{-3(\gamma-1)} \phi^{(\beta+1)}   =0,
\label{hamilton-jacobi}
\end{equation}
with $\rm \eta_\gamma=192\pi a_0 M_\gamma$. In the following we obtain classical solutions for any value in the
 barotropic parameter $\rm p=\gamma\rho$ using the Lagrange-Sharpit approach \cite{Elsgoltz,delgado,lopez}.

\subsubsection{Lagrange-Sharpit approach}

For apply this method, we begin using the equation (\ref{hamilton-jacobi}), where we introduce the following transformations
\begin{eqnarray}
\rm \phi=e^{\Phi}, \qquad A=e^\Omega,\label{trans}
\end{eqnarray}
into (\ref{hamilton-jacobi}), we have
\begin{equation}
\rm -6 \left(\frac{\partial S}{\partial\Phi}\right)^2 + \omega
\left(\frac{\partial S}{\partial\Omega}\right)^2
 +6 \left(\frac{\partial S}{\partial\Omega}\right) \left(\frac{\partial S}{\partial\Phi}\right)
-\eta_\gamma  e^{-3(\gamma-1)\Omega}  e^{(\beta+1)\Phi}   =0.\,
\label{hamilton-jacobi1-2}
\end{equation}

We start defining a function $\rm F(\Phi,\Omega,p,q,S)=0$, where
\begin{equation}
\rm F(\Phi,\Omega,p,q,S)=-6p^2+\omega q^2+6pq  -\eta_\gamma
e^{-3(\gamma-1)\Omega}e^{(\beta+1)\Phi} ,\,\label{F}
\end{equation}
with the expressions $p=\frac{\partial S}{\partial\Phi}$ and
$q=\frac{\partial S}{\partial \Omega}$. We build the following set of
equations
\begin{equation}
 \rm
 \frac{d\Phi}{F_p}=\frac{d\Omega}{dq}=\frac{dS}{pF_p+qF_q}=
  -\frac{dp}{F_\Phi+pF_S}=
  - \frac{dq}{F_\Omega+qF_S}=dt.\label{set}
\end{equation}

calculating the derivatives we have
\begin{eqnarray*}
\rm F_p&=&\rm -12p + 6q,\qquad F_q= 2\omega q + 6p, \qquad F_\Phi= -
(\beta+1) \eta_\gamma   e^{-3(\gamma-1)\Omega}
e^{(\beta+1)\Phi}, \label{Phi}\\
\rm F_\Omega&=& \rm 3 (\gamma-1)\eta_\gamma  e^{-3(\gamma-1)\Omega}
e^{(\beta+1)\Phi},\qquad F_S=0,\qquad F_t=0,\label{s}
\end{eqnarray*}
equation (\ref{set}) is read as
\begin{eqnarray}
\rm  \frac{d\Phi}{-12p+6q}&=&\rm  \frac{d\Omega}{2\omega q + 6 p}= \frac{dS}{p[-12p+6q]+q[2\omega q + 6p]}\nonumber\\
\rm&=& \frac{dq}{-3(\gamma-1)\eta_\gamma
e^{-3(\gamma-1)\Omega}e^{(\beta+1)\Phi}}=\frac{dp}{(\beta+1)\eta_\gamma
e^{-3(\gamma-1)\Omega}e^{(\beta+1)\Phi} }=\rm dt.
\end{eqnarray}
Choosing the equation
\begin{equation}
\rm \frac{dq}{-3(\gamma-1)\eta_\gamma
e^{-3(\gamma-1)\Omega}e^{(\beta+1)\Phi}}
=\frac{dp}{(\beta+1)\eta_\gamma
e^{-3(\gamma-1)\Omega}e^{(\beta+1)\Phi} },
\end{equation}
we find the following relation between the functions (p,q),
\begin{equation}
\rm (\beta +1)q+3(\gamma-1)p=c_\gamma=constant,\label{relation-1}
\end{equation}
with this and from (\ref{F}), substituting the term $\eta_\gamma e^{-3(\gamma-1)\Omega}e^{(\beta+1)\Phi}$, we write an equation which involves to
dependence of time,
$$\rm \frac{(\beta+1)dp}{a_1p^2+a_2p+a_3}=dt,$$
where the solution is
\begin{equation}
\rm p(t)=-\frac{a_2}{2a_1}-\frac{\sqrt{a_2^2-4a_1 a_3}}{2a_1} \,
\coth\left[c_\gamma\sqrt{3a_0}t \right],
\end{equation}
and the constants $a_i$ are
\begin{equation}
\rm a_1=
-6(\beta+1)^2+9\omega(1-\gamma)^2+18(1-\gamma)(\beta+1),\qquad
 a_2=6c_\gamma\left[\omega(1-\gamma)+\beta+1\right],\qquad
 a_3= \omega c_\gamma^2. \nonumber
\end{equation}
From (\ref{relation-1}) we obtain the corresponding solution for the
function $\rm q(t)$ as
\begin{equation}
\rm q(t)= \frac{c_\gamma}{\beta+1}+\frac{3(1-\gamma)}{\beta +1}p(t).
\end{equation}

Now we found the solution for the functions $(p,q)$, then we use the following equation in order to obtain the function $\Phi(t)$,
\begin{equation}
\rm \frac{d\Phi}{-12p+6q}=dt,
\end{equation}
being
\begin{equation}
\rm \Phi(t)=a_4 t+a_5
Ln\left[Sinh(c_\gamma\sqrt{3a_0}t)\right]+\Phi_{0}\,,
 \end{equation}
where $\Phi_{0}$ is an integration constant, and the constants
\begin{equation}
\rm a_4=
\frac{6c_\gamma}{\beta+1}+\frac{3a_2}{\beta+1}\frac{(3\gamma+2\beta-1)}{a_1},\qquad
a_5 =\frac{6(3\gamma+2\beta-1)}{a_{1}} .\nonumber
\end{equation}

For obtain the solution for the function $\Omega$ we use the equation
\begin{equation}
\rm dt= \frac{d\Omega}{2\omega q + 6 p},
\end{equation}
after integration, we have
\begin{equation}
\rm \Omega(t)=a_6\, t -a_7 \,Ln\left[Sinh(c_{0}\sqrt{3a_0}t)\right]+\Omega_{0} \,,
\end{equation}
where $\Omega_{0}$ is an integration constant, and the constants
\begin{equation}
\rm a_6= \frac{1}{\beta +1}\left[2\omega
c_\gamma-\frac{3a_2}{a_1}\left(1+\beta+\omega(1-\gamma)\right)\right],\qquad
 a_7=\frac{6(1+\beta+\omega(1-\gamma))}{a_{1}} .\nonumber
\end{equation}

Now, taking in account the transformation law (\ref{trans}), we have the solution for the scale factor $A$ and the scalar field $\phi$ as
\begin{equation}
\rm A(t)=A_{0}e^{a_6 t} Sinh^{-a_7}(c_\gamma\sqrt{3a_0}t),\qquad
\phi(t)= \phi_{0}e^{a_4 t} Sinh^{a_5}(c_\gamma\sqrt{3a_0}t) \,.
\label{solu:a}
\end{equation}

By introducing these solutions in the equation (\ref{dal}), the
value of the constant $c_\gamma$ is equal to

\begin{equation}\label{constante}
c_\gamma=\sqrt{64\pi a_{1}M_{\gamma}}A_{0}^{\frac{3(1-\gamma)}{2}}\phi_{0}^{\frac{\beta+1}{2}}
\,.
\end{equation}
The corresponding value for all constants that appear in the
classical calculation are presented in table \ref{classi}.
\begin{center}
 \begin{table}[h]
\begin{tabular}{|c|c|c|c|c|}
\hline
& $\gamma=-1$  & $\gamma=\frac{1}{3}$ & $\gamma=1$ & $\gamma=0$ \\
& & & & \\
\hline
$\beta$  &   $2-\lambda a_{0}$  & $0$ & $\frac{\lambda a_{0}-2}{2}$   & $\frac{2-\lambda a_{0}}{4}$ \\
\hline
$a_{1}$  &
$6(9+6\omega-\lambda^{2}a_{0}^{2})$
&  $2a_{0}$
& $-\frac{3}{2}\lambda^{2}a_{0}^{2}$  & $\frac{3}{8}(36+24\omega-\lambda^{2}a_{0}^{2})$ \\
\hline
$a_{4}$ & $\frac{6c_{-1}}{3-\lambda a_{0}}\left[1+(2\omega+3-\lambda a_{0})\ell_0\right]$ &
$6c_{1/3}$
 & $0$ &
  $\frac{3c_{0}}{6-\lambda a_{0}}\left[8+(4\omega+6-\lambda a_{0})\ell_1\right]$ \\
\hline
$a_{5}$  & $2\ell_{0}$    &   $0$  & $\frac{-4}{\lambda a_{0}}$ & $\ell_1$ \\
\hline
$a_{6}$  &
$\frac{c_{-1}}{3-\lambda a_{0}}\left[2\omega-\frac{3(1-\lambda)^{2}a_{0}^{2}}{9+6\omega-\lambda^{2}a_{0}^{2}}\right]$
& $-3c_{1/3}$ &
$\frac{(4\omega-6)c_{1}}{\lambda a_{0}}$ &
 $\frac{4c_{0}}{6-\lambda a_{0}}\left[2\omega-\frac{3(4\omega+6-\lambda a_{0})^{2}}{36+24\omega-\lambda^{2}a_{0}^{2}}\right]$ \\
\hline
$a_{7}$  &
$\frac{(1-\lambda)a_{0}}{9+6\omega-\lambda^{2}a_{0}^{2}}$ &
$1$ & $\frac{-2}{\lambda a_{0}}$ &
$\frac{4(4\omega+6-\lambda a_{0})}{36+24\omega-\lambda^{2}a_{0}^{2}}$ \\
\hline
$c_{\gamma}$  & $\sqrt{384\pi(9+6\omega-\lambda^{2}a_{0}^{2})M_{-1}}A_{0}^{3}\phi_{0}^{\frac{3-\lambda a_{0}}{2}}$ & $\sqrt{128\pi a_{0}M_{1/3}}A_{0}\phi_{0}^{\frac{1}{2}}$ & $\sqrt{-96\pi\lambda^{2}a_{0}^{2}M_{1}}\phi_{0}^{\frac{\lambda a_{0}}{4}}$ & $\sqrt{24\pi(36+24\omega-\lambda^{2}a_{0}^{2})M_{0}}A_{0}^{\frac{3}{2}}\phi_{0}^{\frac{6-\lambda a_{0}}{8}}$ \\
\hline
\end{tabular}
\caption{\label{classi} \small{\emph{ All constants that appear in the
classical solutions for various values of $\gamma$, with the following definitions $\rm
\ell_0=\frac{-\lambda a_{0}}{9+6\omega-\lambda^{2}a_{0}^{2}}$,
$\rm \ell_1=\frac{-8\lambda a_{0}}{36+24\omega-\lambda^{2}a_{0}^{2}}$}}}
\end{table}
\end{center}
\normalsize
When we calculate the deceleration parameter
\begin{equation*}
q=-\frac{A\ddot{A}}{\dot{A}^{2}}
\end{equation*}
where the sign of q indicated whether the model inflates or not. The
positive sign of q i.e. $\rm q>0$ correspond to standard
decelerating model, whereas the negative sign $\rm q < 0$ indicates
acelerate expansion.

Using (\ref{solu:a}), this parameter have the following form
\begin{equation}\label{q}
q=-1-\frac{3a_{7}a_{0}c_{\gamma}^{2}}{\left[\sqrt{3a_{0}}a_{7}c_{\gamma}\cosh\left(c_{\gamma}\sqrt{3a_{0}}t\right)-
a_{6}\sinh\left(c_{\gamma}\sqrt{3a_{0}}t\right) \right]^{2}} \,,
\end{equation}
and checking the table \ref{classi} over the possible value to the
constants $\rm a_0, a_7$ and $\rm c_\gamma$, we observed that when
$\omega>\frac{7}{2}$, this deceleration parameter always is negative
for any value in the barotropic parameter $\gamma$, so, the universe
expand always in this theory.

\subsection{Quantum scheme}
Imposing the quantization condition and applying the Hamiltonian density (\ref{hami1}) to the wave function $\Psi$, we obtain the
 WDW equation for these models  in the minisuperspace  by the usual identification $\rm P_{q^\mu}= -i \partial_{q^\mu}$
 in (\ref {hami1}), with $\rm q^\mu=(A,\phi)$, and following Hartle and Hawking {\cite {HaHa}}
  we consider a semi-general factor ordering which gives (we include the equation (\ref{rho}) in this last equation)
  (Also we could use the loop quantum cosmology in this approach
  \cite{ya1,ya2})
\begin{equation}
\rm \hat{H}\Psi=\left[6\phi^2 \frac{\partial^2}{\partial \phi^2}- 6r
\phi\frac{\partial}{\partial\phi}-\omega
A^2\frac{\partial^2}{\partial A^2}
 +\omega Q A \frac{\partial}{\partial A}
- 6 A \frac{\partial}{\partial A} \phi \frac{\partial}{\partial
\phi}-\eta_\gamma A^{-3(\gamma-1)} \phi^{\beta+1} \right]\Psi=0,
\label{w}
\end{equation}
where Q and r are real constants that measures the ambiguity in the
factor ordering between the scalar functions $\rm (A,\phi)$ and its
corresponding momenta.

Using the same transformation as the classical scheme, (\ref{trans}), the equation (\ref{w}) is read as
\begin{equation}
 \rm \left[6 \frac{\partial^2}{\partial \Phi^2}-
6(r+1)\frac{\partial}{\partial\Phi}-\omega
\frac{\partial^2}{\partial\Omega^2}
 +\omega (Q+1)\frac{\partial}{\partial \Omega}
- 6 \frac{\partial^2}{\partial \Omega \partial \Phi}-\eta_\gamma
e^{-3(\gamma-1)\Omega} e^{(\beta+1)\Phi} \right]\Psi=0, \label{w-m}
\end{equation}
re-written this equation in the following form,
\begin{equation}
\rm 6 \rm \frac{\partial^2 \Psi}{\partial \Phi^2}- 6
\frac{\partial^2\Psi}{\partial \Omega \partial \Phi}-\omega
\frac{\partial^2 \Psi}{\partial \Omega^2}=
 -\omega (Q+1)  \frac{\partial\Psi}{\partial \Omega}
 +\eta_\gamma e^{-3(\gamma-1)\Omega} e^{(\beta+1)\Phi} \Psi+ 6(r+1)
\frac{\partial \Psi}{\partial\Phi}, \nonumber
\end{equation}
we can build the following $\sigma$ parameter
\begin{equation}
\rm \sigma^{\pm}=-\frac{1}{2}\pm \frac{\sqrt{3a_0}}{6},
\label{parameter}
\end{equation}
which is used to construct a linear partial differential equation of first order of the characteristics surfaces for the functions $(\Phi,\Omega)$, given by
\begin{equation}
\rm \frac{\partial \theta}{\partial \Phi}+\sigma^{\pm} \frac{\partial
\theta}{\partial \Omega}=0,\nonumber\\
\end{equation}
where $\theta$ is the envelope of the characteristic surfaces, then we have
$$\rm d\Phi=\frac{d\Omega}{\sigma^{\pm}}=\frac{d\theta}{0},$$
once we integrate this result gives us the following relation
between these functions $\rm \sigma^{\pm} \Phi-\Omega=b_0$.

Let $\rm \Omega=\sigma^{\pm}  \Phi -b_0$, then introducing this in the equation (\ref{w-m}), we have
\begin{equation*}
 \rm 6 \frac{\partial^2 \Psi}{\partial \Phi^2}- 6(r+1)
\frac{\partial \Psi}{\partial\Phi}-\omega \frac{\partial^2
\Psi}{\partial \Omega^2}
 +\omega (Q+1)  \frac{\partial \Psi}{\partial \Omega}
  - 6 \frac{\partial^2 \Psi}{\partial \Omega \partial \Phi}
 -\eta_\gamma e^{3(\gamma-1)b_0} e^{[\beta+1 -3(\gamma-1)\sigma^{\pm}]\Phi}
 \Psi=0.
\label{w-mm}
 \end{equation*}
Assuming that the wavefunction is separable $\rm
\Psi(\Omega,\Phi)=\psi_1(\Omega) \psi_2(\Phi)$ we have
\begin{equation}
 \rm \frac{6}{\psi_2} \frac{d^2 \psi_2}{d \Phi^2}-
\frac{6(r+1)}{\psi_2} \frac{d\psi_2}{d\Phi}-\frac{\omega}{\psi_1}
\frac{d^2 \psi_1}{d \Omega^2}
 +\frac{\omega (Q+1)}{\psi_1}  \frac{d \psi_1}{d \Omega}
  - 6\frac{1}{\psi_1} \frac{d \psi_1}{d \Omega}\frac{1}{\psi_2}\frac{d \psi_2}{d \Phi}
 -\eta_\gamma e^{3(\gamma-1)b_0} e^{[\beta+1 -3(\gamma-1)\sigma^{\pm}]\Phi}
 =0,\nonumber
\end{equation}
and considering the particular ansatz for solving this equation, $\rm \frac{1}{\psi_1}\frac{d^2 \psi_1}{d\Omega^2}=\mu_0^2$, which solution become $\rm\psi_1=b_1\,e^{-\mu_0 \Omega}$, and $\rm b_1$ is a real constant. We choose only the minus sign, because we need to keep a decreasing wavefunction depending to scale factor. With this, the last equation is written as
\begin{equation}
 \rm \frac{d^2 \psi_2}{d \Phi^2}+ b_2  \frac{d
 \psi_2}{d\Phi} - (b_3 + b_4 e^{b_5 \Phi})\psi_2=0,\label{wdw1}
 \end{equation}
 where the constants are defined as
 \begin{equation}
 \rm b_2= \mu_0 -1-r,\qquad b_3= \frac{\omega  \mu_0}{6}\left[  \mu_0+1+
 Q\right],\qquad
 b_4= \frac{\eta_\gamma}{6} e^{3(\gamma-1)b_0}, \qquad
b_5=\beta+1 -3(\gamma-1)\sigma^{\pm}, \nonumber
\end{equation}
 equation that is similar to \cite{polyanin}
 \begin{equation}
 \rm y^{\prime\prime}+ a  y^\prime +(b e^{nx} + c)y=0,
 \end{equation}
 which solution is

 $$ \rm y=e^{\frac{-ax}{2}}\left[c_1 J_\nu\left(\frac{2}{n}\sqrt{b} e^{\frac{nx}{2}} \right)
 + c_2 Y_\nu\left(\frac{2}{n}\sqrt{b} e^{\frac{nx}{2}} \right)\right], $$
 where $\nu=\frac{1}{n}\sqrt{a^2 -4c}$, having the following relations between the parameters:
 $$x=\Phi,\quad a=b_2,\quad b=-b_4,\quad
 n=b_5,\quad c=-b_3,$$
 then substituting this relations and using the transformation equation $\phi=e^\Phi$, we have the following solution to $\psi_2$
 \begin{equation}
\rm \psi_2=\phi^{\frac{1+r-\mu_0}{2}}\left[c_1 K_\nu\left(\kappa
\phi^{\frac{b_5}{2}} \right)
 + b_2 I_\nu\left(\kappa \phi^{\frac{b_5}{2}} \right)\right], \qquad
 \nu=\frac{1}{b_5} \sqrt{\xi},
 \end{equation}
 where $\rm (K_\nu,I_\nu)$ are the modified Bessel functions, $\kappa=\frac{2\sqrt{\eta_\gamma/6}}{b_5}e^{\frac{3(\gamma-1)b_0}{2}}$
 and  $\xi=(1+r-\mu_0)^2+\frac{2\omega  \mu_0}{3}\left[\mu_0+1+Q\right]$.
 Then, the wavefunction have the following form
\small{
\begin{equation}\label{wavefuncton}
\rm \Psi_\nu= A^{-\mu_0}\,\phi^{\frac{1+r-\mu_0}{2}}\left[C_1
K_\nu\left(\kappa \phi^{\frac{b_5}{2}} \right)+C_2 I_\nu\left(\kappa
\phi^{\frac{b_5}{2}} \right)\right].
\end{equation}}
The corresponding value for all constants that appear in the quantum
calculation are presented in table \ref{quantum}.
\begin{center}
\begin{table}[h]
\begin{tabular}{|c|c|c|c|c|}
\hline
$\gamma$ &     $0$      &    $\frac{1}{3}$   & $1$ & $-1$ \\
\hline $b_{5}$  &
$\frac{6-\lambda a_{0}}{4}+3\sigma^{\pm}$
&
 $1+2\sigma^{\pm}$ & $\lambda a_{0}$ & $3-\lambda a_{0}+6\sigma^{\pm}$ \\
\hline $\nu$    & $\frac{1}{b_5}\sqrt{\xi}$ &
$\frac{1}{1+2\sigma^{\pm}}\sqrt{\xi}$ &
$\frac{1}{\lambda a_{0}}\sqrt{\xi}$ &
 $\frac{1}{b_5}\sqrt{\xi}$ \\
\hline $\kappa$ &
$\frac{2}{b_{5}}\sqrt{\frac{\eta_{0}}{6}}e^{-\frac{3}{2}b_{0}}$
         & $\frac{2}{1+2\sigma^{\pm}}\sqrt{\frac{\eta_{1/3}}{6}}e^{-b_{0}}$ & $\frac{2}{\lambda a_{0}}\sqrt{\frac{\eta_{1}}{6}}$ &
          $\frac{2}{b_{5}}\sqrt{\frac{\eta_{-1}}{6}}e^{-3b_{0}}$ \\
\hline
\end{tabular}
\caption{ \label{quantum} \emph{All constants that appear in the
quantum solutions for various values of $\gamma$. Here
$\xi=(1+r-\mu_0)^2+\frac{2\omega \mu_0}{3}\left[\mu_0+1+Q\right]$}}.
\end{table}
\end{center}

\section{Conclusions}
In this paper we have investigated the flat FRW cosmological model of
the universe in the framework of Barber's second self-creation
 theory since of point to view of hamiltonian systems, where the classical solutions were found under the Hamilton-Jacobi
 approach combined with the Lagrange-Charpit method for all values of the barotropic parameter in a perfect fluid.
 By means of the deceleration parameter we obtain that the universe
 in this theory always suffers an expansion in any epoch of the
 universe.
 In the quantum behavior, the solutions
were found using the curves characteristics method applied to
partial differential equation of second degree, obtaining the
solutions, as in the classical scheme, for all values in the
$\gamma$ parameter.

\acknowledgments{This work was supported in part by  DAIP
(2011-2012), Promep  UGTO-CA-3 and CONACyT 167335, 179881  grants.
JMR was supported by Promep grant ITESJOCO-001. Many calculations
where done by Symbolic Program REDUCE 3.8. This work is part of the
collaboration within the Advanced Institute of Cosmology, and Red
PROMEP: Gravitation and Mathematical Physics under project {\it
Quantum aspects of gravity in cosmological models, phenomenology and
geometry of space-time}.}\\

 \end{document}